\documentclass{svjour2}
\smartqed
\usepackage{graphicx}
\usepackage{amsmath}

\begin{document}
\title{Classical and Quantum Mechanics via Supermetrics in Time}
\titlerunning{Supermetrics in Time}
\author{E. GOZZI}
\institute{Department of Theoretical Physics, University of Trieste \\
Strada Costiera 11, Miramare-Grignano 34014, Trieste\\ and INFN,  
Sezione di Trieste, Italy\\
e-mail: gozzi@ts.infn.it or ennio.gozzi@gmail.com
}
\maketitle

\begin{abstract}
 Koopman-von Neumann in the 30's gave an operatorial formululation of Classical Mechancs.
 It was shown later on that this formulation  could  also be written in a path-integral form.  We will label  this functional approach as CPI (for classical path-integral) to distinguish it from the quantum mechanical one, which we will indicate with QPI.  In the CPI two Grassmannian partners of time make their natural appearance and in this manner time becomes something like a three dimensional supermanifold. Next we introduce a metric in this supermanifold and show that a particular choice of the supermetric reproduces the CPI while  a different one gives the QPI. 

\end{abstract}

\section{INTRODUCTION.}
\label{intro}
The topic of this conference has been Spin-Statistics.  One of the things which is most difficult to accept to our common sense is the anticommuting nature of Fermions. Actually anticommuting variables have a long history that goes back to Grassmann \cite {gras} and are not strictly related to spin. Grassmann invented them for abstract reasons but  he discovered that they are  usefull in the description of ruled surfaces. Later on the exterior product introduced by Grassmann was used in the field of differential forms by Cartan. Grassmannian variables have made their appearance in theoretical physics thanks to the work of Faddeev-Popov who realized that they can  represent some "ghost" degrees of freedom in gauge theory  and are necessary for the unitarity of the theory.(for a review  see for example \cite{fap} ) In the 70's and 80's the Russian School , centered on the late Berezin \cite{ber}, greatly developed  the topic of Grassmannian variables inventing an integral and differential calculus  for them. Besides "gauge ghosts" they have been used in theoretical physics to give a "classical" descripton of spin\cite{casal} and to build the first point-particle Lagrangian exhibiting both global\cite{witten} and local \cite{divecchia} supersymmetry.

In this paper we want to bring forward a further use of Grassmannian quantities which made their appearance once we  give a path-integral formulation to Classical Mechanics\cite{ennio}. This formulation is nothing else than the functional counterpart of the Koopman-von Neumann operatorial approach to classical mechanics\cite{koop} and has nothing to do with quantum mechanics. To distinguish the two we will label one as CPI and the other as QPI.

In the CPI, besides the configurational , or phase space, variables $x$, some auxiliary variables made their appearances: they are two anticommuting  $c,{\bar c}$ and one commuting  $\lambda$ variables. Moreover time $ t $ naturally comes equipped with two Grassmannian  partners $\theta,{\bar\theta}$
and together they make what is commonly known as  super-time. The variable $x$ and its partners, via the use of the super-time, can be put together in a "multiplet" called superfield:
\begin{displaymath}
X(t,\theta,\bar{\theta})=x(t)+\theta c(t)+\bar{\theta}\bar{c}+i\bar{\theta}\theta \lambda(t)
\end{displaymath}
The interesting thing that we have shown \cite{gq} is that the CPI  has the same functional form as the QPI. The only difference is that we have to replace the fields $x(t)$ of
the QPI with the superfields
and the integrations over the time variable $t$ with integrations over the supertime
 $(t,\theta,\bar{\theta})$. The quantization can then be realized\cite{gq} by sending to zero
 the two Grassmann partners of time $\theta$ and $\bar{\theta}$.
 
 In these notes we would like to investigate the possibility of realizing the quantization of the system
 in a ``more natural" way. By ``more natural" we mean that we would not like to
 send ``by hand" the $\theta,\bar{\theta}$ to zero in order to pass from the CPI
 to the QPI. We would like to achieve this transition in a more ``geometrical" manner.
 It is true that looking at this transition as a sort of ``dimensional reduction"\cite{ghir}, the whole thing 
 is already geometrical in its nature but in most theories people talk about dimensional
 reduction only w.r.t. ``bosonic" coordinates while here we have Grassmannian coordinates.
 In this paper  we will build a more general path integral than the CPI or the QPI.
 We will construct a single generating functional which is invariant under general
 super-diffeomorphisms in $(t,\theta,\bar{\theta})$. This is possible by introducing a
 suitable supermetric or supervierbein in the space $(t,\theta,\bar{\theta})$. Equipped
 with this formalism we shall then show that a particular set of metrics gives the weight
 of the CPI and another one reproduces the weight of the QPI. Unfortunately the metric
 associated to the QPI is a non-invertible one. It is anyhow possible to introduce a
 unified formalism via a ``regularized metric" which reproduce the CPI for some value
 of the regularizing paramater and the QPI for some other value. We will provide below
 some further details of this unified formalism whose first details appeared in ref.\cite{ghir}

\section{METRIC AND VIERBEIN.}
\label{sec:1}

Since we want to implement a new path integral, invariant under general coordinate
 transformations in the supertime we are very close to the spirit of the first papers
 on supergravity\cite{divecchia}\cite{arnow}.Consequently we can use some of their structures and results.
 In supergravity people mostly use\cite{arnow} the supervierbeins $E$ instead of the supermetrics $g$.
 Then, modulo grading factors, one can reconstruct the metric from the vierbein via the
 following equation
\begin{equation}
g_{\scriptscriptstyle MN}=E^{\scriptscriptstyle A}_{\scriptscriptstyle M}
\eta_{\scriptscriptstyle AB} (-1)^{\scriptscriptstyle
(1+B)N}E^{\scriptscriptstyle B}_{\scriptscriptstyle N}, \label{metvier}
\end{equation}
where $\eta_{\scriptscriptstyle AB}$ is the analogous of the Minkowski metric.
In our approach the ``space-time" or base space is made up by one even variable $t$
and two odd variables $\theta$ and $\bar{\theta}$. The Minkowski metric will be  symmetric
in the even-even part and antisymmetric in the odd-odd part. We can choose for example:
\begin{displaymath}
\displaystyle \eta_{\scriptscriptstyle{AB}}=\begin{pmatrix}
1 & 0 & 0 \cr 0 & 0 &-1 \cr 0 & 1 & 0
\end{pmatrix}.
\end{displaymath}
The analogue of the Lorentz group is given by the orthosymplectic group OSp(1,2)
whose generators are:
\begin{eqnarray}
&& \displaystyle X_{\scriptscriptstyle 1}=-\bar{\theta}\partial_{\theta}, \qquad
X_{\scriptscriptstyle 2}=\theta \partial_{\bar{\theta}}, \qquad X_{\scriptscriptstyle 3}
=-\frac{1}{2}(\bar{\theta}\partial_{\bar{\theta}}-\theta\partial_{\theta}) \nonumber \\
&& \displaystyle X_{\scriptscriptstyle 4}=-\frac{1}{\sqrt{2}}(\bar{\theta}
\partial_t-t\partial_{\theta}), \qquad X_{\scriptscriptstyle 5}=\frac{1}{\sqrt{2}}
(\theta\partial_t+t\partial_{\bar{\theta}}) \label{osp12}.
\end{eqnarray}
The operators $X_i$, with $i=1,\cdots 5$, of Eq. (\ref{osp12}) leave invariant the
distance of a point of the supertime $z^{\scriptscriptstyle A}
\equiv (t,\theta,\bar{\theta})$ from the origin, i.e. the following quantity:
\begin{displaymath}
\displaystyle F\equiv z^{\scriptscriptstyle A} \eta_{\scriptscriptstyle AB}
z^{\scriptscriptstyle B}=t^2-2\bar{\theta}\theta.
\end{displaymath}
Flat distances in supertime , like the one above, had first been introduced in ref.\cite{ucraini}
and further details have been worked out in appendix B of \cite{gq}.

Before going on, let us review some mathematical formulae\cite{dewitt} that we will use
in the following sections. First of all, the superdeterminant of a supermatrix
 $\begin{pmatrix} A & B \cr C & D \end{pmatrix}$ is defined by de Witt as follows:
\begin{displaymath}
\displaystyle \textrm{sdet} \begin{pmatrix} A & B \cr C & D \end{pmatrix}
\equiv \textrm{det}(A-BD^{-1}C) \textrm{det}^{-1} D,
\end{displaymath}
where $A, D$ are Grassmannian odd blocks and $B, C$ are Grassmannian even blocks.
The inverse of a supermatrix is given by:
\begin{displaymath}
\displaystyle \begin{pmatrix} A & B \cr C & D \end{pmatrix}^{-1}=
\begin{pmatrix}
(1-A^{-1}BD^{-1}C)^{-1}A^{-1} & -(1-A^{-1}BD^{-1}C)^{-1}A^{-1}BD^{-1}
\cr -(1-D^{-1}CA^{-1}B)^{-1}D^{-1}CA^{-1} & (1-D^{-1}CA^{-1}B)^{-1}D^{-1}.
\end{pmatrix}
\end{displaymath}
In a generic Grassmann number:
\begin{displaymath}
\displaystyle f(\theta, \bar{\theta})=f_{\scriptscriptstyle 0}+f_{\scriptscriptstyle 1}
\theta
+f_{\scriptscriptstyle 2}\bar{\theta}+f_{\scriptscriptstyle 3}\theta \bar{\theta}
\end{displaymath}
it is called ``body" \cite{dewitt}. that part of the number which does not depend on $\theta$ or $\bar{\theta}$.
The remaining part is called \cite{dewitt} the ``soul" of the Grassmann number 
Its inverse is given by:
\begin{displaymath}
\displaystyle f^{-1}(\theta,\bar{\theta})=\frac{1}{f_{\scriptscriptstyle 0}}
-\frac{f_{\scriptscriptstyle 1}}{f_{\scriptscriptstyle 0}^2}\theta
-\frac{f_{\scriptscriptstyle 2}}{f_{\scriptscriptstyle 0}^2}\bar{\theta}
-\frac{f_{\scriptscriptstyle 3}}{f_{\scriptscriptstyle 0}^2}
\theta \bar{\theta}.
\end{displaymath}
It is clear from this fact that the inverse exists only when
$f_{\scriptscriptstyle 0}\neq 0$. This is equivalent to say that the body
of the Grassmann number $f$ must be different than zero.

\section{INVARIANT ACTION AND THE CPI.}
\label{sec:2}

An action invariant under generic reparametrizations of the supertime
$(t,\theta, \bar{\theta})$ can be build in analogy to what people have done
in supergravity\cite{divecchia}
\begin{equation}
\displaystyle S=i\int \textrm{d}t\textrm{d}\theta \textrm{d}\bar{\theta} \,
E \left[ \frac{1}{2}D_tX(t,\theta,\bar{\theta})D_tX(t,\theta,\bar{\theta})-V(X)\right],
\label{action}
\end{equation}
where $D_t\equiv E^{\scriptscriptstyle{M}}_t\partial_{\scriptscriptstyle{M}}$
and $E\equiv\textrm{sdet}(E^{\scriptscriptstyle{A}}_{\scriptscriptstyle{M}})$
are constructed from the matrix of the  vierbeins  and its inverse: $E^{\scriptscriptstyle{M}}
_{\scriptscriptstyle{A}}$ and $E^{\scriptscriptstyle{A}}_{\scriptscriptstyle{M}}$.
In order to check for the invariance of the action above, we choose the superfield $X(t,\theta,
\bar{\theta})$ to
transform as a scalar under general coordinate transformations. From Eq.
(\ref{action}) we see that, to reproduce the right potential term $V(X)$ of the
CPI \cite{gq} , it is crucial to impose the condition $E=1$. Moreover
the right kinetic terms of the CPI \cite{gq} are obtained  if the following equation holds:
\begin{displaymath}
\displaystyle D_tXD_tX=\partial_tX\partial_tX.
\end{displaymath}
Let us parametrize the vierbein matrix $E^{\scriptscriptstyle M}_{\scriptscriptstyle A}$
as follows:
\begin{equation}
\displaystyle E^{\scriptscriptstyle{M}}_{\scriptscriptstyle{A}}\equiv
\begin{pmatrix}
a & \alpha & \beta \cr
\gamma & b & c \cr
\delta & d & e  \label{vier}
\end{pmatrix}
\end{equation}
where $a, b, c, d$ and $e$ are Grassmannian even and $\alpha, \beta, \gamma$ and
$\delta$ are Grassmannian odd. In terms of the entries of the metric
$E^{\scriptscriptstyle{M}}_{\scriptscriptstyle{A}}$ the covariant derivative
$D_t$ which appears in the action (\ref{action}) can be expressed as:
\begin{displaymath}
\displaystyle D_t\equiv E^{\scriptscriptstyle M}_t\partial_{\scriptscriptstyle M}
=a\partial_t +\alpha \partial_{\theta}+\beta\partial_{\bar{\theta}}.
\end{displaymath}
Consequently the kinetic term becomes:
\begin{displaymath}
D_tXD_tX=a^2\partial_t X\partial_t X+2a\alpha \partial_t
X\partial_{\theta}X+2a\beta \partial_tX\partial_{\bar{\theta}}X
\end{displaymath}
and reproduces the right kinetic term $\partial_tX \partial_tX$ of the CPI iff
 $\alpha=\beta=0$ and $a=\pm 1$. So there remain six free entries in the vierbein:
\begin{equation}
\displaystyle E^{\scriptscriptstyle{M}}_{\scriptscriptstyle{A}}\equiv
\begin{pmatrix}
\pm 1 & 0 & 0 \cr
\gamma & b & c \cr
\delta & d & e. \label{matCPI}
\end{pmatrix}
\end{equation}
Let us remember that the above vierbein must satisfy the condition on the determinant,
i.e. $\textrm{sdet}E^{\scriptscriptstyle{M}}_{\scriptscriptstyle{A}}=1$. This implies:
\begin{equation}
\displaystyle \textrm{det} \begin{pmatrix}
 b & c \cr
 d & e
\end{pmatrix}=\pm 1 \, \Rightarrow \, be-cd =\pm 1. \label{det}
\end{equation}
If we rewrite $b, c, d, e$ in terms of their ``bodies" $b_{\scriptscriptstyle B}$ and souls
$b_{\scriptscriptstyle S}$, e.g. $b=b_{\scriptscriptstyle{B}}
+b_{\scriptscriptstyle{S}}\bar{\theta}\theta$, then the condition (\ref{det})
splits in two conditions, one for the body and one for the soul:
\begin{equation}
\left\{ \begin{array}{l}
\displaystyle b_{\scriptscriptstyle{B}}e_{\scriptscriptstyle{B}}-
c_{\scriptscriptstyle{B}}d_{\scriptscriptstyle{B}}=\pm 1  \medskip \\
\displaystyle b_{\scriptscriptstyle{S}}e_{\scriptscriptstyle{B}}+
b_{\scriptscriptstyle{B}}e_{\scriptscriptstyle{S}}-
c_{\scriptscriptstyle{S}}d_{\scriptscriptstyle{B}}-
c_{\scriptscriptstyle{B}}d_{\scriptscriptstyle{S}}=0. \label{4bis}
\end{array}
\right.
\end{equation}
whose solutions are
\begin{displaymath}
\displaystyle b_{\scriptscriptstyle{B}}=\frac{\pm 1+c_{\scriptscriptstyle{B}}
d_{\scriptscriptstyle{B}}}{e_{\scriptscriptstyle{B}}}, \qquad
b_{\scriptscriptstyle{S}}=\frac{\mp e_{\scriptscriptstyle{S}}-
c_{\scriptscriptstyle{B}}d_{\scriptscriptstyle{B}}e_{\scriptscriptstyle{S}}+
c_{\scriptscriptstyle{B}}d_{\scriptscriptstyle{S}}e_{\scriptscriptstyle{B}}+
c_{\scriptscriptstyle{S}}d_{\scriptscriptstyle{B}}e_{\scriptscriptstyle{B}}}
{e_{\scriptscriptstyle{B}}^2} \quad \textrm{with} \quad
e_{\scriptscriptstyle{B}}\neq 0
\end{displaymath}
or
\begin{displaymath}
\displaystyle c_{\scriptscriptstyle{B}}=\mp\frac{1}{d_{\scriptscriptstyle{B}}},
\qquad
c_{\scriptscriptstyle{S}}=\frac{\pm d_{\scriptscriptstyle{S}}+
b_{\scriptscriptstyle{B}}d_{\scriptscriptstyle{B}}e_{\scriptscriptstyle{S}}}
{d_{\scriptscriptstyle{B}}^2} \quad \textrm{with} \quad e_{\scriptscriptstyle{B}}= 0.
\end{displaymath}
If we pass from the vierbeins to the metrics using Eq. (\ref{metvier}), the freedom in
the choice of a metric compatible with the CPI can be parametrized by just five
variables, as it emerges from the following expression:
\begin{equation}
g_{\scriptscriptstyle MN}=\begin{pmatrix}
1 & \mp \pi_1 \theta \mp \pi_2\bar{\theta} &
\mp \pi_3\theta \mp \pi_4\bar{\theta} \cr
\pm \pi_1\theta\pm \pi_2\bar{\theta} & 0 & \mp 1
\mp\pi_5 \bar{\theta}\theta \cr
\pm \pi_3\theta \pm \pi_4\bar{\theta} & \pm 1 \pm
\pi_5\bar{\theta}\theta & 0
\end{pmatrix}
\end{equation}
where we have defined the following new parameters:
\begin{equation}
\left\{ \begin{array}{l}
\pi_1 \equiv \gamma_{\theta}e_{\scriptscriptstyle B}-\delta_{\theta}
c_{\scriptscriptstyle B}  \medskip \\
\pi_2 \equiv \gamma_{\bar{\theta}}e_{\scriptscriptstyle B}-
\delta_{\bar{\theta}}c_{\scriptscriptstyle B} \medskip \\
\pi_3 \equiv \delta_{\theta}b_{\scriptscriptstyle B}-
\gamma_{\theta}d_{\scriptscriptstyle B}  \medskip \\
\pi_4 \equiv \delta_{\bar{\theta}}b_{\scriptscriptstyle B}-
\gamma_{\bar{\theta}}d_{\scriptscriptstyle B} \medskip  \\
\pi_5 \equiv \gamma_{\bar{\theta}}\delta_{\theta}-
\gamma_{\theta}\delta_{\bar{\theta}}.
\end{array} \right.  \label{pi}
\end{equation}
If we use the formulae given in \cite{arnow}
we can calculate the dependence of the Ricci scalar from the parameters $\pi_i$.
In particular, the body of the Ricci scalar turns out to be :
\begin{displaymath}
\displaystyle R_{\scriptscriptstyle B}=-\frac{1}{2}\pi_{2}^2
-\frac{1}{2}\pi_{3}^2+ 5\pi_{2}\pi_{3}
-\pi_{2}^2\pi_{5}-\pi_{3}^2\pi_{5}-2\pi_{2}\pi_{\scriptscriptstyle 3}
\pi_{5}-4\pi_{1}\pi_{4}+4\pi_{1}\pi_{4}\pi_{5}\pm 6\pi_{5}.
\end{displaymath}
From this expression we can say that, due to the freedom we have in
$\pi_5$ (and, most probably also in the $\pi_1,\pi_2,\pi_3,\pi_4$ of Eq.
(\ref{pi})) every value of the body of the Ricci scalar is compatible with
the metrics of the CPI. This implies that  the various metrics belonging to the set 
which reproduces the CPI action (that means those satisfying the constraints above)
 are not necessarily diffeomorphic equivalent.

\section{ON THE KINETIC TERM.}
\label{sec:3}

Of course the previous parametrization depends on the particular kinetic term
that we have chosen $D_tXD_tX$. Are there other kinetic terms invariant
under general coordinate transformations and compatible with the weight of the CPI?
In particular, let us consider the following action:
\begin{equation}
\displaystyle S=i\int \textrm{d}t\textrm{d}\theta \textrm{d}\bar{\theta} \,
 E \left[ \frac{1}{2}D_{\theta}XD_{\bar{\theta}}X-V(X)\right], \label{action2}
\end{equation}
where $D_{\theta}\equiv E^{\scriptscriptstyle{M}}_{\theta}
\partial_{\scriptscriptstyle{M}}$
and $D_{\bar{\theta}}\equiv E^{\scriptscriptstyle{M}}_
{\bar{\theta}}\partial_{\scriptscriptstyle{M}}$. Assuming again
that $X$ transforms as a scalar, the action ({\ref{action2}}),
like the one in (\ref{action}), is invariant under general
coordinate transformations of the supertime $(t,\theta, \bar{\theta})$.
Nevertheless, there is no choice of the vierbein
$E^{\scriptscriptstyle M}_{\scriptscriptstyle A}$ which allow us to
reproduce the weight of the CPI.
In fact, to reproduce the potential, we must choose, as in the previous section,
$E=1$ while to reproduce the right kinetic term we must impose that:
\begin{displaymath}
\displaystyle D_{\theta}XD_{\bar{\theta}}X=\partial_tX\partial_tX.
\end{displaymath}
In terms of the entries of the matrix (\ref{vier}) we can write:
\begin{eqnarray}
\displaystyle D_{\theta}XD_{\bar{\theta}}X &=& \displaystyle
\gamma\delta (\partial_t X)^2+ (
\gamma d-\delta b)\partial_tX\partial_{\theta}X+
(\gamma e -\delta c)\partial_t
X\partial_{\bar{\theta}}X+(cd-be)
\partial_{\bar{\theta}}X\partial_{\theta}X \nonumber \\
&=& \partial_tX \partial_tX
\end{eqnarray}
from which we get the following conditions:
\begin{equation}
\left\{ \begin{array}{l}
\gamma\delta=1, \qquad \quad \gamma d-\delta b=0  \\
\gamma e -\delta c=0, \qquad cd-be =0. \label{cond1}
 \end{array} \right.
\end{equation}
The condition $E=1$ gives:
\begin{eqnarray}
\displaystyle \textrm{sdet}\begin{pmatrix}
a & \alpha & \beta \cr
\gamma & b & c \cr
\delta & d & e
\end{pmatrix}&=& \frac{a}{be-dc}-\frac{1}{(be-dc)^2}\begin{pmatrix} \alpha &\beta
\end{pmatrix} \begin{pmatrix} e & -c \cr -d & b \end{pmatrix}
\begin{pmatrix}
\gamma \cr \delta \end{pmatrix} \nonumber \\
&=& \frac{a}{be-dc}-\frac{1}{(be-dc)^2}\left[\alpha(e\gamma-c\delta)-
\beta(d \gamma-b \delta)\right]=1. \nonumber
\end{eqnarray}
Using (\ref{cond1}) the square brackets in the equation above is identically
zero and so the condition on the superdeterminant reduces to:
\begin{displaymath}
\frac{a}{be-dc}=1
\end{displaymath}
which is incompatible with the last of Eq. (\ref{cond1}).

\section{INVARIANT ACTION AND THE QPI.}
\label{sec:4}

Let us start from the action (\ref{action}) and let us see if we can reproduce
also the weight of the QPI by choosing different vierbeins. In order to
achieve the dimensional reduction \cite{gq} which brings the weight (\ref{action})
into the quantum one \cite{gq} which has a potential of the form $V(x)$ and not of the form  $V(X)$, we have to choose as superdeterminant the following one:
\begin{equation}
\displaystyle E=-\frac{i\bar{\theta}\theta}{\hbar}. \label{sdet2}
\end{equation}
This creates a technical difficulty. In fact a Grassmann number with zero body,
like the one of Eq. (\ref{sdet2}), does not admit an inverse. Let us bypass
this problem by introducing a small ``regularizing" parameter $\epsilon$ and
define a ``regularized" superdeterminant as:
\begin{equation}
\displaystyle E=\epsilon -\frac{i\bar{\theta}\theta}{\hbar}. \label{sdet3}
\end{equation}
The expression (\ref{sdet2}) and the weight of the QPI can be obtained
in the limit $\epsilon \to 0$. The inverse of $E$ in (\ref{sdet3}) has the
following expression:
\begin{equation}
E^{-1}=\frac{1}{\epsilon}+\frac{i\bar{\theta}\theta}{\epsilon^2\hbar}. \label{detqu}
\end{equation}
This will be the determinant of the vierbein
$E^{\scriptscriptstyle M}_{\scriptscriptstyle A}$ which enter the definition
of the kinetic terms of the action (\ref{action}). Such an action becomes:
\begin{equation}
\displaystyle S=i\epsilon
\int \textrm{d}t \textrm{d}\theta \textrm{d}\bar{\theta}\,
\left[\frac{1}{2}D_tXD_tX-V(X)\right]
+\frac{1}{\hbar}\int \textrm{d}t\textrm{d}\theta
\textrm{d}\bar{\theta} \, \bar{\theta}\theta
\left[\frac{1}{2}D_tXD_tX-V(X)\right]. \label{quant}
\end{equation}
The first term goes to zero when $\epsilon \to 0$ while the second term,
performing the Grassmann integrations in $\theta, \bar{\theta}$, reduces to
\begin{displaymath}
\frac{1}{\hbar} \int \textrm{d}t \, \left[ \frac{1}{2}
a_{\scriptscriptstyle B}^2 \partial_t x \partial_t x-V(x) \right].
\end{displaymath}
This is the weight of the QPI, provided $a_{\scriptscriptstyle B}=\pm 1$.
This is the only condition coming from the kinetic terms. The other two
conditions are related to the body and the soul of the superdeterminant of (\ref{vier}):
\begin{equation}
E^{-1}=\left(\pm 1 +a_{\scriptscriptstyle S}\bar{\theta}
\theta-p\bar{\theta}\theta\right)
(q+r\bar{\theta}\theta) = \pm q+(a_{\scriptscriptstyle S}q-
pq\pm r)\bar{\theta}\theta,
\label{detqua}
\end{equation}
where we have defined:
\begin{equation}
p\bar{\theta}\theta \equiv \begin{pmatrix} \alpha &\beta
\end{pmatrix} \begin{pmatrix} b & c \cr d & e \end{pmatrix}^{-1}
\begin{pmatrix}
\gamma \cr \delta \end{pmatrix}, \qquad
q+r\bar{\theta}\theta \equiv \textrm{det}^{-1}
\begin{pmatrix} b & c \cr d & e \end{pmatrix}.
\label{defpi}
\end{equation}
When we equate (\ref{detqu}) and (\ref{detqua}) we get that the
following two constraint equations:
\begin{equation}
q=\pm \frac{1}{\epsilon}, \qquad \pm
\left(\frac{a_{\scriptscriptstyle S}}{\epsilon}-
\frac{p}{\epsilon}+r\right)=\frac{i}{\epsilon^2\hbar}. \label{constr}
\end{equation}
The second of Eq. (\ref{defpi}) implies that the matrix
$\displaystyle D\equiv \begin{pmatrix} b & c \cr d & e \end{pmatrix}$
must be invertible and the determinant of its inverse be equal to
\begin{displaymath}
q+r\bar{\theta}\theta=\pm \frac{1}{\epsilon}+r\bar{\theta}{\theta}.
\end{displaymath} So $D$ must have determinant
$\pm \epsilon -\epsilon^2r \bar{\theta}\theta$ which implies:
\begin{displaymath}
\displaystyle be-cd=\pm \epsilon -\epsilon^2 r \bar{\theta}\theta.
\end{displaymath}
This is equivalent to the following system of equations:
\begin{eqnarray}
&& \displaystyle b_{\scriptscriptstyle B}e_{\scriptscriptstyle B}-
c_{\scriptscriptstyle B}d_{\scriptscriptstyle B}=\pm \epsilon \nonumber\\
&& \displaystyle b_{\scriptscriptstyle S}e_{\scriptscriptstyle B}
+b_{\scriptscriptstyle B}e_{\scriptscriptstyle S}
-c_{\scriptscriptstyle S}d_{\scriptscriptstyle B}
-c_{\scriptscriptstyle B}d_{\scriptscriptstyle S}
=-\epsilon^2 r.  \label{one}
\end{eqnarray}
The first equation is a constraint equation, the second equation allows us to
write $r$ as a function of the entries of the vierbein. In the same way, if we
split the Grassmann odd entries of the vierbein as
$\alpha=\alpha_{\theta}\theta+\alpha_{\bar{\theta}}\bar{\theta}$
and we use the definition of $p$ coming from (\ref{defpi}) we can write
also $p$ as a function of the entries of the vierbein. Replacing such an expression
and (\ref{one}) into (\ref{constr}) we get another constraint equation between the
entries of the vierbein.
The constraint equations can be solved but the solution is not very enlightening.

As a particular case, let us put $a_{\scriptscriptstyle S}=\alpha=\beta=0$,
as in the classical solution. In such a case $p=0$ and the constraint equations
reduce to the following ones:
\begin{equation}
\left\{ \begin{array}{l}
\displaystyle b_{\scriptscriptstyle{B}}e_{\scriptscriptstyle{B}}-
c_{\scriptscriptstyle{B}}d_{\scriptscriptstyle{B}}=\mp \frac{i}
{\hbar}  \medskip \\
\displaystyle b_{\scriptscriptstyle{S}}e_{\scriptscriptstyle{B}}+
b_{\scriptscriptstyle{B}}e_{\scriptscriptstyle{S}}-
c_{\scriptscriptstyle{S}}d_{\scriptscriptstyle{B}}-
c_{\scriptscriptstyle{B}}d_{\scriptscriptstyle{S}}=
\pm \epsilon. \label{15}
\end{array} \right.
\end{equation}
If we take the following expressions:
\begin{displaymath}
\displaystyle b_{\scriptscriptstyle{B}}=\frac{\pm \epsilon+
c_{\scriptscriptstyle{B}}d_{\scriptscriptstyle{B}}}
{e_{\scriptscriptstyle{B}}}, \qquad
b_{\scriptscriptstyle{S}}=\frac{\mp
\epsilon e_{\scriptscriptstyle{S}}-c_{\scriptscriptstyle{B}}
d_{\scriptscriptstyle{B}}e_{\scriptscriptstyle{S}}+c_{\scriptscriptstyle{B}}
d_{\scriptscriptstyle{S}}e_{\scriptscriptstyle{B}}+c_{\scriptscriptstyle{S}}
d_{\scriptscriptstyle{B}}e_{\scriptscriptstyle{B}}
\mp(1-\epsilon)\frac{i}{\hbar} e_{\scriptscriptstyle B}}{e_{\scriptscriptstyle{B}}^2},
\quad \textrm{with} \quad e_{\scriptscriptstyle B}\neq 0
\end{displaymath}
and
\begin{displaymath}
\displaystyle c_{\scriptscriptstyle{B}}=\mp\frac{\epsilon}
{d_{\scriptscriptstyle{B}}},\qquad
c_{\scriptscriptstyle{S}}=\frac{\pm \epsilon d_{\scriptscriptstyle{S}}
+b_{\scriptscriptstyle{B}}d_{\scriptscriptstyle{B}}
e_{\scriptscriptstyle{S}}\pm (1-\epsilon)\frac{i}{\hbar}
d_{\scriptscriptstyle B}}{d_{\scriptscriptstyle{B}}^2},
\quad \textrm{with} \quad e_{\scriptscriptstyle B}= 0.
\end{displaymath}
we can reproduce the weight of the QPI in the limit $\epsilon \to 0$,
but we can reproduce also the weight of the CPI in the limit $\epsilon \to 1$.

\section{OPEN PROBLEMS.}
\label{sec:5}

We like to list here some problems we hope to attack in the future:

\subsection{}The supervierbeins which satisfy the constraint (\ref{4bis})
for the CPI and (\ref{one}) for the QPI are not just one but several.
Actually it is possible to count how many free parameters are left in the
supervierbein. For example from Eq. (\ref{matCPI}) we see that we have
12 parameters in the case of the CPI minus the two constraint equations 
of (\ref{4bis}), i.e. 10 free parameters. Those sets of vierbeins leave
the weight of the CPI invariant so, if we could connect them
 by a continuous transformation, they would make  a Noether symmetry of the system.
 This symmetry is not necessary a superdiffeomorphism. In fact those vierbeins,
 and the associated metrics, have only the determinant (1 for the CPI and
 $-i \bar{\theta}\theta/\hbar$ for the QPI) in common which is not in general
 an invariant of the diffeomorphisms. We cannot even say that the symmetry
 is the set of unimodular diffeomorphisms which are those which preserve
 the determinant. In fact our metrics or vierbeins are a subset of those
 with the same determinant; they are that particular subset which produces a kinetic
 piece of the form $\partial_tX\partial_tX$. 
So the first problem is: are the set of metrics which gives  the right determinant and the right 
kinetic term connected by a
continuous transformation? Or is there a manner to parametrize the solutions
of (\ref{4bis}) and (\ref{15}) in order to get a better grasp of the
``geometrical meaning" of these solutions inside the CPI (or inside QPI) set.
A manner to get a meaning would be to understand geometrically what it means
a kinetic term of the form $\partial_tX\partial_tX$ (and for the QPI the corresponding one\cite{gq}).

\subsection{}
 The parametrization of the supermetrics in the QPI case should be studied more in details. In particular it should be understood why we have a different amount of freedom in the choice of the metrics for the CPI  with respect to the QPI.
Moreover, like for the CPI, we should check if the metrics of the QPI gives different curvatures. If that is most probably the case, this means that also the set of metrics reproducing the QPI are not diffeomorphic equivalent to one another.

\subsection{} The next problem is the following. Let us take the local invariance of our action
seriously, and let us  make a diffeomorphism starting from any metric inside the CPI set.
We will  get a set of equivalent theories which do not respect the CPI  constraints but, nevertheless, they are equivalent (in the superdiffeomorphic sense) to Classical Mechanics. The same for the QPI. The question now is: if someone gives us  a Lagrangian in the enlarged space, is there a
quick manner to find out if this  is a  system ``diffeomorphic  equivalent" 
to a CPI  or to  a QPI one ?
That means :"Can we calculate some quantities from that Lagrangian which  tell us 
if that system is equivalent to a classical or to a quantum one ?"
These ``quantities" would really be the ``true signature" of the classical (
or quantum) nature of the system. They are somehow like the "moduli" of the trajectories obtained by a differeomorphism which starts from a point inside the CPI (or QPI) set.

\subsection{} A further  problem would be to find out the ``Einstein equation" for our
supervierbein (or supermetric) i.e., something analog to
\begin{equation}
\displaystyle G^{\mu \nu}=k T^{\mu \nu}. \label{3-1}
\end{equation}
The metrics of the CPI could  be a solution of this equation when a particular
``matter content" is chosen which produces a particular $T^{\mu \nu}$. The same
for the QPI. Of course the $T^{\mu \nu}$ cannot be given only by the matter fields
$X$ because these are the same both for the CPI and the QPI and they would give
the same $T^{\mu \nu}$. As the CPI metrics are not diffeomorphism equivalent to
the QPI ones (because these ones are not invertible) via a regular diffeomorphism,
they cannot be obtained from the same $T^{\mu\nu}$.  So  there
must be some ``hidden matter" which produces extra pieces in the
$T^{\mu \nu}$ and different  from each other for  the CPI and QPI cases.
 For the Eq.
(\ref{3-1}) one also should find out which is the $k$ constant .

It might also be that the equation of motion for the metric is not an Einstein-like equation ?. Somehow, like in string theory, we do not have anything providing the dynamics of the background metric. A principle analog to the equivalence principle (for time) has to be searched.

\subsection{}  Basically we should ask ourselves  if   our introduction of a metric, to get a unified picture of the QPI and CPI, is just a mathematical artifact or has it a physical meaning? If it is a mathematical
artifact, has it some internal inconsistency? The fact that the solutions inside
 each set  (the CPI and the QPI one) are not diffeomorphism equivalent to each other does
 it mean that they cannot be solutions of the same covariant equation?
 
 \subsection{} As we cannot pass from the QPI  set of metrics to the CPI  ones via a differomorphism, is there any other kind of flow (like the Ricci or the Calabi ones (for a review see for example \cite{calabi}) which can bring us from one to the other ?. The parameter of the flow could be the regularizing parameter $\epsilon$  mentioned above. As most of Ricci and Calabi  flows are related  to renormalization flows, finding  one in our case would give indications that we can pass from the QPI to the CPI via a sort of "renormalization" procedure. Actually, independent of the issue of the supermetric, we have been working \cite{ennios} for a while on techniques similar to renormalization which would bring us from the QPI to the CPI . The approach is the following: start from the QPI  and apply  a sort of block-spin technique, that means in the QPI the phase-space variables are replaced by small cells of phase-space. Once these cells have a size bigger than
the Planck constant, the QPI should turn into the CPI . This is so because, at that regime, we should get Classical Mechanics but in a path-integral form because the QPI, from which we started, was  a path-integral. Note that in the CPI we had   already found sort of different variables than the phase-space points present in the QPI, we had the superfields as basic ingredients to functionally integrate over. These superfields  may be the "mathematical representation" of the block-spin\cite{ennios}. Actually, if we take the QPI as the fundamental theory with its uncertainty principle and the functional integration over the phase-space, then in Classical Mechanics we cannot use the phase-space points as basic variables because of the uncertainty principle. We have to use  the blocks of phase-space of size
bigger than the Planck constant or equivalently the superfields. These are the real degrees of freedom at the classical level. The symmetries that we found for the CPI\cite{ennio} are basically the freedom we have in parametrizing (or moving  into)  the block-spins. Freedom that we loose once we shrink the cells to a point like it happens in the QPI. Let us now  consider the fact that the action of the QPI and the CPI formally remains the same as was  proved in \cite{gq}  and that  we have only to replace the fields with the superfields. This  is equivalent to saying that only the fields get "renormalized". This is a typical feature of supersymmetric theories and ours \cite{ergo} is one. The associated beta function of this "renormalization" flow should be related to the Ricci o Calabi flow we were looking for . Finally we should study  the zeros of this beta function, especially the ultraviolet stable ones and the analog  infrared ones.  They may tell us if Classical Mechanics and Quantum Mechanics are the ultimate theories at respectively the largest and smallest scales\cite{ennios}

\subsection{}
Last but not least , as it is a supertime, and not time, the central object  that makes its appearance in this study, we should try to throw further light on the  physical meaning of the Grassmannian  partners of time. A first study \cite{ennios} seems to indicate that they are related to the square roots of infinitesimal intervals of time, someting like fluctuations of time but more work is needed.
 
\section{CONCLUSION.}
Despite the technical details that the reader had to go through in this paper, we feel that some intuitive physical conclusions can be summarized here. Basically the message we get is that Classical Mechanics   and Quantum Mechanis measures the intervals of time in different manners.
Both in the CPI and the QPI we have a lot of freedom in chosing these time intervals, because both the CPI set and the QPI one are wide enough but for sure, by reparametrizing time, we cannot bring a time interval measured using "Classical Tools" into one obtained by using "Quantum Tools".
So  the differences in the time intervals seem to be what is at the heart of the differences between Classical and Quantum Mechanics. We have been able to indentify this issue of time as the central feature  because we have managed, via the path-integral, to make all the other features of Classical Mechanics  and Quantum mechanics  as similar as possible to each other. 

Somehow the message that we got  from this analysis is "similar" to the one  we got from the black-body radiation problem: at small frequencies (long intervals of time) we should use Classical Mechanics, while at high frequencies (short intervals of time) we should use Quantum Mechanics. Of course this analysis is rather naive and intuitive but it may be what is really behind the mathematical work presented in this paper.

\begin{acknowledgements}

{This paper would not have been written scientifically without the help and great generosity of my good friend Danilo Mauro who shared with me, over a period of 5 years, many calculations contained in this paper. It was finally written up and  in Latex thanks to the help  of Ale,Carlo and Gianna. This work has been supported by INFN  and MIT via a Bruno Rossi exchange fellowship with MIT. The hospitality , support  and exchange of ideas with various faculty members and students at MIT is gratefully acknowledged.}
\end{acknowledgements}


\begin{thebibliography}{}

\bibitem{gras} Grassmann H.,Die lineare ausdehnungslehre, Wiegand, Lipzig (1844)
\bibitem{fap} Jan Govaerts, Hamiltonian quantisation and constrained dynamics, Leuven University Press, Leuven, (1991) 
\bibitem{ber} Berezin F.A., Introduction to superanalysis, Reidel Publishing Company, Dordrecht ,(1987)
\bibitem{casal} Casalbuoni R, On the quantization of systems with anticommuting variables, Nuovo Cimento A 33, 115 (1976); Berezin F.A.  and  Marinov M.S, Particle spin dynamics as the Grassmann variant of classical mechanics, Ann. Phys. (N.Y.) 104, 336  (1977)
\bibitem{witten} Witten E,Dynamical breaking of supersymmetry, Nucl.Phys.B188 , 513 (1981)
\bibitem{divecchia}Brink  L. et al.,Local supersymmetry for spinning particles, Phys.Lett. B64, 435 (1976)
\bibitem{ennio}Gozzi E., Reuter M., Thacker W.D., Hidden BRS invariance in classical mechanics:II, Phys.Rev.D 40 , 3363 (1989)
\bibitem{koop} Koopman B.O., Hamiltonian systems and transformations in Hilbert space, Proc. Nat. Acad. Sci. USA 17, 315 (1931); von Neumann J., Zur operatorenmethode in der klassichen mechanik, Ann. Math. 33, 587  (1932)
\bibitem{gq}Abrikosov A.A. Jr., Gozzi E. and Mauro D., Geometric dequantization, Ann.Phys (N.Y.) 317, 24 (2005)
\bibitem{ghir}Gozzi, E. and  Mauro D.,Quantization as a dimensional reduction phenomenon , AIP Conf.Proc.84 ,158, (2006)
\bibitem{arnow}Arnowitt  R. and  Nath P., Riemannian Geometry in spaces with Grassmann coordinates , Gen.Rel. Grav 7, 89 (1976)
\bibitem{ucraini}Tugai V.V. and Zheltukhin, A.A., Superfield generalization of the classical action-at- a-distance theory,  Phys.Rev.D 51, 3997 (1995)
\bibitem{dewitt}De Witt B. Supermanifolds, Cambridge University Press, Cambridge, (1987)
\bibitem{calabi}  Bakas I.: Geometric Flows and (some of) their physical applications, hep-th-0511057
\bibitem{ennios} Gozzi E. et al. : Work in progress. 
\bibitem{ergo} Gozzi E.,Reuter, M. Algebraic characterization of ergodicity, Phys.Lett.B 233 ,3, 383 (1989);E.238,2, 451 (1990)

\end{thebibliography}
\end{document}